\documentstyle[12pt,epsfig]{article}   
\def\Journal#1#2#3#4{{#1} {\bf #2}, #3 (#4)}

\def\NPB{{\em Nucl. Phys.} {\bf B}}
\def\PLB{{\em Phys. Lett.} {\bf B}}
\def\PRL{\em Phys. Rev. Lett.}
\def\PRD{{\em Phys. Rev.} {\bf D}}
\def\ZPC{{\em Z. Phys.} {\bf C}}
\def\EPJ{{\em Eur. Phys. J.} {\bf C}}
\def\lsim{\mathrel{\rlap{\lower4pt\hbox{\hskip1pt$\sim$}}
    \raise1pt\hbox{$<$}}}                
\def\gsim{\mathrel{\rlap{\lower4pt\hbox{\hskip1pt$\sim$}}
    \raise1pt\hbox{$>$}}}                

\def\ra{\rightarrow}
\def\be{\begin{equation}}
\def\ee{\end{equation}}
\def\bea{\begin{eqnarray}}
\def\eea{\end{eqnarray}}
\newcommand{\SC}{\langle{\cal O}_8^{\psi '}(^1S_0)\rangle}

\newcommand{\SDsing}{\langle{\cal O}_1^{\psi '}(^3S_1)\rangle}
\newcommand{\PS}{\langle{\cal O}_8^{\psi '}(^3P_0)\rangle}

\begin{document}
\begin{flushright}
KOBE--FHD--99--05\\
December~~~~~~~1999
\end{flushright}
\vspace*{0.5cm}
\begin{center}
\baselineskip=1cm
{\Large \bf Two--Spin Asymmetry for {\boldmath $\psi^{\prime}$} 
Photoproduction \\with Color--Octet Mechanism} \\
\baselineskip=0.6cm
\vspace{2cm}
{\large K.~Sudoh\footnote[2]{E-mail: sudou@radix.h.kobe-u.ac.jp} 
and T.~Morii\footnote[1]{E-mail: morii@kobe-u.ac.jp}} \\
\vspace{0.5cm}
{\em Faculty of Human Development and Graduate School of Science and 
Technology,  \\ Kobe University, Nada, kobe, 657--8501, Japan}
\vspace{4cm} \\
\end{center}

\noindent
\begin{center}
{\bf Abstract}
\end{center}

We studied the photoproduction of $\psi^{\prime}$ in the forward regions in 
polarized $\gamma p$ collisions at relevant HERA energies. 
We found that this reaction is very effective to test the color--octet 
mechanism which is based on the NRQCD factorization formalism. 
Furthermore we found that the value of the NRQCD matrix elements can be 
severely constrained by measuring the two--spin asymmetry, 
though the process depends on the polarized gluon distribution $\Delta 
g(x)$. 

\clearpage

Heavy quarkonium productions and decays have been traditionaly calculated 
in the framework of the color--singlet model \cite{Schuler94}. 
However, it has been reported that the color--singlet model cannot 
explain the experimental data of $J/\psi$ and $\psi^{\prime}$ 
hadroproductions \cite{CDF92}; the cross sections of prompt $J/\psi$ and 
$\psi^{\prime}$ production in unpolarized $p\bar{p}$ collisions predicted by 
the color--singlet model were smaller than the Tevatron data by more than 
one order of magnitude \cite{Cho96}. 
Furthermore, in the photoproduction of $J/\psi$, the color--singlet 
model cannot satisfactorily reproduce cross sections for inclusive $J/\psi$ 
production at the recent HERA experiment \cite{H199} even including 
next--to--leading order corrections \cite{Kramer96}. 
In order to solve these serious problems, a new color--octet model has been 
advocated by several people \cite{Braaten95} as one of the most promising 
candidates that can possibly remove such big discrepancies between the 
experimental data and the theoretical prediction by the color--singlet model. 

A rigorous formulation of the color--octet model has been introduced based 
on an effective field theory called nonrelativistic QCD (NRQCD), 
in which the ${\cal{O}}(v)$ corrections of a relative velocity 
between the bound heavy quarks can be systematically calculated 
\cite{Bodwin95}. 
The NRQCD factorization approach separates the effects of short distances 
that are comparable to or smaller than the inverse of heavy quark mass, 
from the effects of longer distance scales of hadronization. 
A heavy $Q\bar{Q}$ pair is first produced in a virtual color--octet 
intermediate state of the NRQCD higher Fock state, and then 
hadronizes into a detected color--singlet particle via the emission or 
absorption of dynamical gluons. 
Production cross sections of heavy quarkonium $H$ can be factorized into 
a product of a short distance coefficient $C_n$ which can be computed using 
perturbative QCD, and a long distance part 
$\langle {\cal O}_{n}^{H}(^{2S+1}L_{J})\rangle$ which is described by 
nonperturbative NRQCD matrix elements whose values should be determined 
from experiments or lattice gauge theory, 
\be
\sum_X d\sigma(AB\ra H+X)=\frac{1}{\Phi}d\Gamma \sum_n C_n 
\langle {\cal O}_{n}^{H}(^{2S+1}L_{J})\rangle ,
\label{eq:NRQCD}
\ee
where $\Phi$ and $\Gamma$ show a flux and a phase space factor, respectively. 
The label $n$ denotes color and angular momentum numbers. 
In this factorization approach, there are two long distance parameters 
which are essentially the vacuum expectation values of the color--singlet 
and --octet NRQCD matrix elements $\langle {\cal O}_{1,8}^H\rangle$, 
whose relative importance is determined by the velocity scaling 
rules \cite{Lepage92}. 
Physics of the color--octet model is now one of the most interesting topics 
on heavy quarkonium productions at high energy. 
However, although the color--octet model is quite succesful in explaining 
the Tevatron data, it looks problematic for the process $\gamma + p \ra 
J/\psi + X$; using the long distance matrix elements extracted from the 
Tevatron data, the color--octet model largely overestimates the HERA 
data \cite{Cacciari96}. 
Since the color--singlet matrix elements is related to the radial 
wave functions at the origin, their values are calculable using 
potential models \cite{Eichten95}. 
On the other hand, the color--octet matrix elements can be extracted 
only from experiments at present. 
However, the present uncertainties on the color--octet matrix elements 
are still large and the discussion seems controversial. 
In order to confirm the validity of the NRQCD factorization approach, 
the universality of long distance parameters must be established for 
different processes with acceptable experimental error. 
So far, several processes have been already suggested for testing this 
factorization approach, such as polar angle distributions of $J/\psi$ 
in $e^+ e^-$ annihilation into $J/\psi + X$ \cite{Braaten96}, $Z^0$ decays at 
LEP \cite{Cheung96}, leptoproduction of $J/\psi$ \cite{Fleming98}, and so on. 
\noindent
\begin{figure}[t]
\begin{center}
\hspace*{0.1cm}
  \epsfxsize=10cm
  \epsfbox{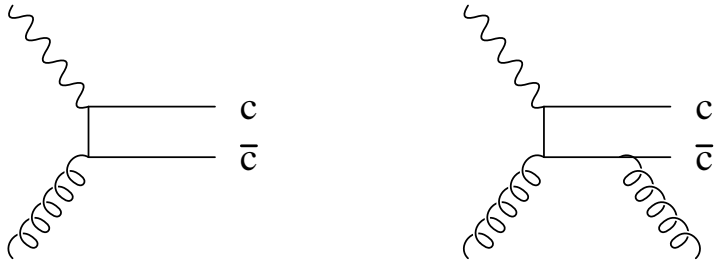}
\end{center}
\vspace{-0.4cm}
\caption[junk]{\it Feynman diagrams for the leading order of the color--octet 
subprocess $\gamma + p \ra (c\bar{c})_{octet}$ and of the color--singlet 
subprocess $\gamma + p \ra (c\bar{c})_{singlet} + g$. 
Initial $\gamma$ and proton are polarized. 
\label{fig:feyndiagram}}
\end{figure}

In this paper, as another test of the color--octet model, we propose the  
$\psi^{\prime}$ photoproduction at small--$p_{T}$ regions, 
\be
\vec{\gamma} + \vec{p} \ra \psi^{\prime} + X , 
\label{eq:process}
\ee
which might be observed in the forthcoming polarized HERA experiment, 
where the arrow attached to particles means that these 
particles are polarized in a parallel or antiparallel direction to the 
running direction of each particle. 
In this work, we do not consider the contribution of elastic diffractive 
mechanisms and other reducible background processes, since they can be 
eliminated with appropriate cuts \cite{Jung92}. 
If the color--octet mechanism works, the process is dominated by the 
color--octet subprocess $\gamma + g \ra (c\bar c)_{octet}$ at 
small--$p_{T}$ regions. 
In addition, there can be the contribution from the conventional 
color--singlet subprocess $\gamma + g \ra (c\bar c)_{singlet} + g$ which 
appears only as a second order of the strong coupling constant $\alpha_s$. 
Feynman diagrams for these subprocesses are illustrated in 
Fig. \ref{fig:feyndiagram}. 
The main contribution to the leading color--octet mechanism comes from 
the color--octet $^{1}S_0^{(8)}$ and $^{3}P_{J=0,2}^{(8)}$ states, 
whereas the color--singlet mechanism originates from $^{3}S_1^{(1)}$ state. 
The process involves the NRQCD long distance parameters corresponding to 
such states. 
In particular, for the color--octet machanism the cross section is 
propotional only to the linear combination of the NRQCD long distance 
parameters, $\SC$ and $\PS$, owing to the NRQCD spin symmetry relation, 
as described later. 
We show that to measure the spin--dependent cross section and two--spin 
asymmetry for this process is very effective not only for testing of the 
color--octet model but also for extracting the values of the color--octet 
long distance matrix elements. 
Furthermore, since the process is dominantly produced in photon--gluon 
fusion, the cross section must be sensitive to the gluon density in a proton 
and thus one can get good information on the spin--dependent gluon 
distribution function by analyzing this polarized process. 
A related subject has been recently investigated by Japaridze {\it et al.} 
\cite{Japaridze99}. They have calculated the two--spin asymmetry of 
$J/\psi$ photoproduction at large--$p_{T}$ regions in polarized 
$\gamma p$ scattering and discussed the sensitivity of the long distance 
parameters to the asymmetry. 
They have insisted that the process is effective for testing the 
color--octet model. 
\noindent
\begin{figure}[t]
\begin{center}
\hspace*{0.1cm}
    \epsfxsize=10cm
    \epsfbox{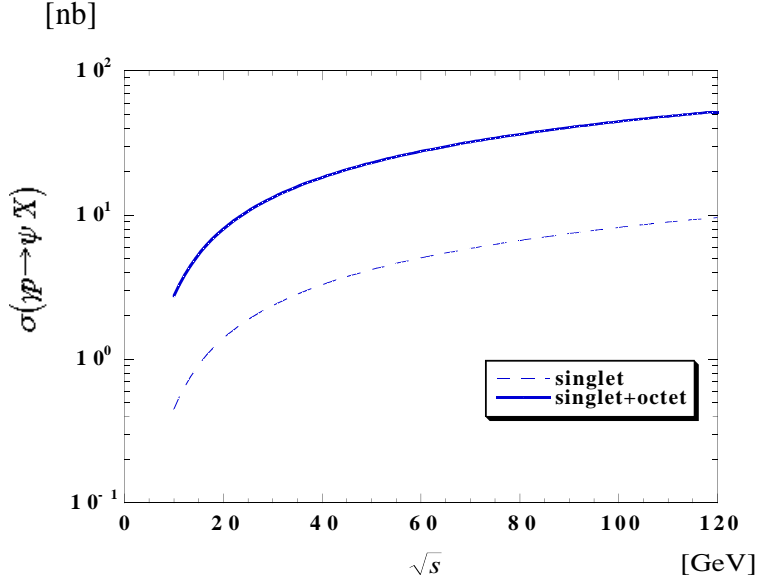}
\end{center}
\vspace*{-0.4cm}
\caption{\it The spin--independent total cross section as a function of 
$\sqrt{s}$. The solid line represents the sum of the color--sinlet and octet 
contributions, while the dashed line represents the color--singlet 
contribution only. 
\label{fig:cross}}
\end{figure}

The spin--independent total cross section due to these mechanisms for 
small--$p_{T}$ regions is given by \cite{Fleming96}
\bea
\sigma(\gamma p \ra \psi^{\prime} + X) 
&=& \sigma_{\underline{8}}(\gamma p \ra \psi^{\prime} + X) + 
\sigma_{\underline{1}}(\gamma p \ra \psi^{\prime} + X) \nonumber \\
&=& \int dx g(x) \left(\sum \hat\sigma(\gamma g \ra c\bar{c} 
\left[^{2S+1}L_{J}^{(8)} \right] )\langle{\cal O}_8^{\psi^{\prime}}
(^{2S+1}L_{J})\rangle \right. \nonumber \\ 
& & \hspace*{5cm} \left. + \hat\sigma(\gamma g \ra c\bar{c} 
\left[^{3}S_{1}^{(1)} \right]g )\langle{\cal O}_1^{\psi^{\prime}}
(^{3}S_{1})\rangle \right) \nonumber \\
&=& \frac{\pi^2 \alpha_em e_c^2}{m_c^3} \int dx g(x) \left(
\pi\alpha_s \delta(4m_c^2 -\hat{s})\Theta \right. \nonumber \\
&+& \left. \int d\hat{t}\frac{64 \alpha_s^2 m_c^4 |R_{\psi^{\prime}}(0)|^2} 
{3\hat{s}^2} 
\frac{\hat{s}^2 (\hat{s}-4m_c^2)^2 + 
\hat{t}^2 (\hat{t}-4m_c^2)^2 + \hat{u}^2 (\hat{u}-4m_c^2)^2}
{(\hat{s}-4m_c^2)^2 (\hat{t}-4m_c^2)^2 (\hat{u}-4m_c^2)^2}\right), 
\nonumber \\ &&
\eea
where $g(x)$ is the unpolarized gluon distribution function, and $\hat{s}$, 
$\hat{t}$ and $\hat{u}$ are the usual Mandelstam variables for the subprocess. 
The labels $\underline{1}$ and $\underline{8}$ for the cross sections denote 
the contribution from the color--singlet and --octet state, respectively. 
The sum in the first term is taken over $^3P_{0,2}^{(8)}$ and $^1S_0^{(8)}$ 
states and $\Theta$ is the linear combination of color--octet matrix elements 
for these states, 
\be
\Theta\equiv\SC +\frac{7}{m_c^2}\PS .
\ee
$R_{\psi^{\prime}}(0)$ is a radial wave function at the origin and is related 
to the color--singlet matrix element as 
\be
\SDsing\sim\frac{9}{2\pi}|R_{\psi^{\prime}}(0)|^2 
\left(1+{\cal O}(v^2) \right) , 
\ee
which can be well determined from the leptonic decay width of 
$\psi^{\prime}$, though its value cannot be calculated perturbatively. 

The spin--dependent cross section can be obtained by replacing the 
unpolarized subprocess cross section by the polarized one, and furthermore 
by the following replacement, 
\bea
&& g(x) \ra \Delta g(x) ~~~~~~~{\rm (\Delta~means~''polarized''.)} , \\
&& \Theta \ra \tilde{\Theta} \equiv \SC - \frac{1}{m_c^2}\PS .
\eea
\noindent
\begin{figure}[t]
\begin{center}
\hspace*{0.1cm}
    \epsfxsize=10cm
    \epsfbox{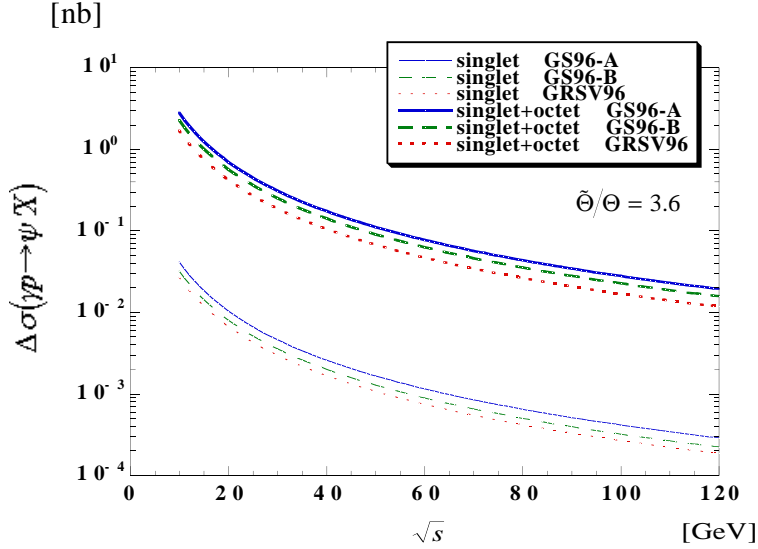}
\end{center}
\vspace{-0.4cm}
\caption[junk]{\it The spin--dependent total cross section with the 
parameter $\tilde{\Theta}/\Theta = 3.6$ as a function of $\sqrt{s}$. 
The solid, dashed and dotted lines show the case of set A of GS96 \cite{GS96}, 
set B of GS96 \cite{GS96} and the 'standard scenario' of GRSV96 \cite{GRSV96}, 
respectively. 
Upper bold lines represent the sum of the color--singlet and --octet 
contributions, while lower lines represent the color--singlet contribution 
only. 
\label{fig:dcross}}
\end{figure}

In the present calculation, we used the GS96(set A and set B) \cite{GS96} and 
GRSV96 \cite{GRSV96} parametrizations for the polarized gluon distribution 
and the GRV95 parametrization \cite{GRV95} for the unpolarized one. 
Color--octet matrix elements were taken from the recent analysis on 
charmonium hadroproduction data: 
$\SC + (7/m_{c}^{2})\PS \approx 5.2\times 10^{-3} {\rm [GeV^{3}]}$ 
\cite{Beneke96} and
$\frac{1}{3}\SC + (1/m_{c}^{2})\PS \approx (5.9\pm 1.9)\times 10^{-3}
{\rm [GeV^{3}]}$ \cite{Leibovich97},
which lead to
\be
\frac{\tilde{\Theta}}{\Theta} \equiv \frac{\SC-\frac{1}{m_{c}^{2}}\PS}
{\SC+\frac{7}{m_{c}^{2}}\PS} \approx 3.6 \sim 8.0 .
\label{eq:ratio}
\ee 
Calculated cross sections are presented in Fig. 
\ref{fig:cross} and Fig. \ref{fig:dcross}. 
We see that the color--octet contribution is larger than the color--singlet 
one by one and two order of magnitude for the unpolarized and polarized 
cross sections, respectively. 
It is remarkable that the difference due to different gluon polarization 
models is not so large. 

Now we move to the analysis on a two--spin asymmetry for $\psi^{\prime}$ 
production in the polarized reaction defined by 
\bea
A_{LL} &\equiv& \frac{\left[d\sigma_{++}-d\sigma_{+-}+
d\sigma_{--}-d\sigma_{-+}\right]}
{\left[d\sigma_{++}+d\sigma_{+-}+
d\sigma_{--}+d\sigma_{-+}\right]} \nonumber \\ 
&=& \frac{d\Delta\sigma}{d\sigma} 
= \frac{d\Delta\sigma_{\underline{8}} + d\Delta\sigma_{\underline{1}} }
{d\sigma_{\underline{8}} + d\sigma_{\underline{1}} } , 
\label{eq:A_LL} 
\eea
where $d\sigma_{+-}$, for instance, denotes that the helicity of photon is 
positive and the one of proton is negative. 
From Eq. \ref{eq:A_LL}, we see that if only the color--octet contribution 
which is the lowest order process in $\alpha_s$ works, the asymmetry 
$A_{LL}$ can be written by a simple formula 
\be
A_{LL}^{\psi^{\prime}} (\gamma p)_{lowest} 
= \frac{d\Delta\sigma_{\underline{8}}}{d\sigma_{\underline{8}}}  
= \frac{\Delta g(x,Q^{2})}{g(x,Q^{2})}\cdot \frac{\tilde{\Theta}}{\Theta} ,
\label{eq:A_LL2}
\ee
which is just a product of the ratio of polarized and unpolarized gluon 
densities and the one of color--octet matrix elements. 

\begin{figure}[t]
\begin{center}
\hspace*{0.1cm}
    \epsfxsize=10cm
    \epsfbox{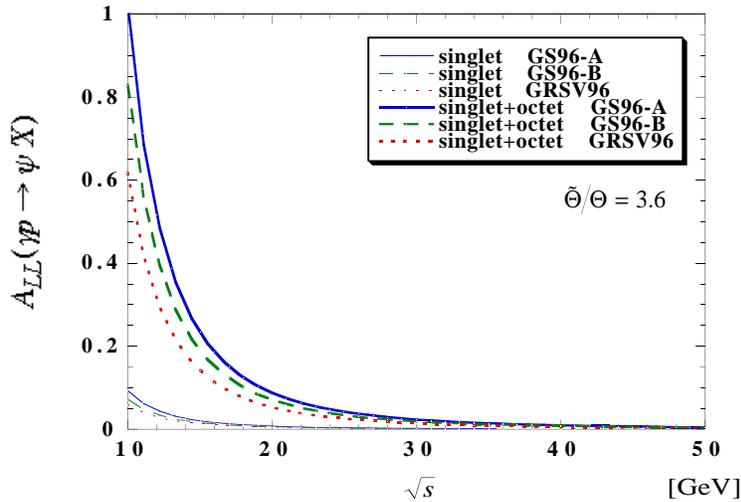}
\end{center}
\vspace{-0.4cm}
\caption{\it The two--spin asymmetry $A_{LL}(\gamma p \ra 
\psi^{\prime}X)$ with the parameter $\tilde{\Theta}/\Theta = 3.6$ as a 
function of $\sqrt{s}$. 
Various lines show the same as in Fig. \ref{fig:dcross}. 
Upper bold lines represent the color--singlet plus octet contribution, 
while lower lines represent the color--singlet contributions only. 
\label{fig:all}}
\end{figure}

Taking $Q^{2}=4m_{c}^{2}$ with a charm mass $m_{c}=1.5 {\rm GeV}$, 
the calculated $A_{LL}$ at relevant HERA energies are presented in Fig. 
\ref{fig:all}. 
As shown in Fig. \ref{fig:all}, if the color--octet process works, then 
the $A_{LL}$ becomes quite large in the rather smaller $\sqrt{s}$ region, 
comparing with the one for the color--singlet mechanism only. 
The difference of $A_{LL}$ due to the color--octet and --singlet mechanism 
is larger than the uncertainties due to the polarized gluon distribution 
functions. 
Hence we can sufficiently test the color--octet contribution in this reaction. 
Furthermore, since the $A_{LL}$ strongly depends on the value of 
$\tilde{\Theta}/\Theta$, one can constrain its magnitude from the value of 
$A_{LL}$ as follows; if we take the GS96 or GRSV96 parametrization which 
are widely used, the maximum value of $\Delta g(x)/g(x)$ becomes roughly 
0.35 for GS96 and 0.2 for GRSV96 and then, with this value on the ratio of 
gluon distributions, we can constrain the maximum value of the ratio of 
NRQCD matrix elements as 
\bea 
\frac{\tilde{\Theta}}{\Theta}~ &\lsim&~ 5.0~~~{\rm for~GS96} ,\\
&\lsim&~ 2.9~~~{\rm for~GRSV96} ,
\label{eq:max}
\eea
from the requirement that the $A_{LL}$ should be less than 1. 
Actually, the uncertainty of matrix elements seems to be larger than 
that of gluon distribution, because the value of matrix elements obtained 
from the Tevatron data which we used here does not include the 
contributions of higher order QCD corrections. 
In Ref. \cite{Kniehl99}, Kniehl and Kramer have approximately taken into 
account the effect from higher order QCD corrections due to multiple--gluon 
initial state radiation and improved the values of long distance parameters 
for $J/\psi$ meson as smaller. 
They have insisted that it is possible to explain both Tevatron and HERA 
data by using such a 'small' set of long distance parameters. 
In any case we can say that our process is very effective not only for 
testing the NRQCD factorization approach but also for constraining the value 
of long distance parameters, though the result depends on the polarized 
gluon distribution function. 
On the contrary, if the NRQCD factorization approach is confirmed enough 
with the long distance matrix elements with acceptable theoretical and 
experimental uncertainties, we can get good information on the polarized gluon 
distribution function in rather smaller $\sqrt{s}$ regions. 

Finally, let us discuss the sensitivity on our results. 
In order to examine the experimental feasibility of the forthcoming HERA 
experiments, we have estimated the experimental sensitivity of the $A_{LL}$ 
for 100--day experiments at various $\sqrt{s}$ in the manner of 
Nowak \cite{Nowak96}, using the expected data of beam or target polarization 
($P_{B}, P_{T} \sim 70\%$), the integrated luminosity (${\cal{L}\cdot T} \sim 
66 {\rm pb^{-1}}$), and the combined trigger and reconstruction efficiency 
($C \sim 50\%$) together with the value of unpolarized total cross sections. 
As a result we found that the experimental sensitivity $\delta A_{LL}$ 
is order of magnitude $\sim 10^{-3}$, which is very small. 
Hence our predictions are expected to be actually tested in the future 
polarized HERA experiments. 

In summary, to test the color--octet model we have proposed the 
photoproduction of $\psi^{\prime}$ at small--$p_{T}$ regions in polarized 
$\gamma p$ scattering which might be available in the forthcoming polarized 
HERA experiments. 
We have calculated two--spin asymmetry $A_{LL}$ for various parameter 
regions $\tilde{\Theta}/\Theta = 3.6 \sim 8.0$, and found that the $A_{LL}$ 
becomes quite large in the regions $\sqrt{s}=10\sim 20$ GeV. 
Therefore we can sufficiently test the color--octet model in this process. 
In addition, the measurement of $A_{LL}$ is very effective to severely 
constrain the value of NRQCD matrix elements, though it depends on the 
polarized gluon distribution $\Delta g(x)$. 
\noindent

\end{document}